# Production Function of the Mining Sector of Iran[1]

Seyyed Ali Zeytoon Nejad Moosavian
*PhD student in Economics at North Carolina State University*

*Abstract*

The purpose of this study is to estimate the production function of Iran's mining sector, and also examine the structure of production in this sector. Several studies have already been conducted in estimating production functions of various economic sectors; however, less attention has been paid to mining sectors. After investigating the stationarity of variables using augmented Dickey-Fuller and Phillips-Perron tests, this study estimated the production function of the mining sector of Iran under different scenarios using the co-integration method and time series data for 1976-2006. The unrestricted Cobb-Douglas production function in the Tinbergen form provided better results in terms of theoretical foundations of economics, statistics, and econometrics. These results suggest that the structure of the mining sector of Iran is both capital-intensive and labour-intensive. Based on the findings of this study, the elasticity of production with respect to capital and labour have been 0.44 and 0.41, respectively. In addition, the coefficient of time variable, as an indicator of technological progress in the production process, is statistically significant representing a positive effect of technological changes on the output of Iran's mining sector.

## Introduction

Mining sector is considered to be one of the main economic sectors that can serve other economic sectors in many direct and indirect ways. Although, in most cases, it has a small direct share in GDP, it usually has a considerable indirect effect on value added creation. The high diversity and abundance of mineral deposits can bring about a great potential for Iran's economy to grow faster. Furthermore, mining sector can be of great importance in creating employment opportunities and balanced regional economic development as well. Iran is a rich country in terms of mineral resources. In order to take advantage of this enormous potential for economic growth and development, policy makers must pay more attention to this economic sector when designing and implementing economic policies (Zeytoonnejad, 2005).

In principle, mining sector refers to the part of the economy which explores, extracts, and processes ores and mineral resources. Based on this definition, all activities other than the three

---

ones mentioned above relate to "mineral industry", which is a separate economic sector. Based on such a classification, economic activities such as smelting, refining, rolling are all classified under mineral industries sector (i.e. mining-related industries sector). Figure (1) clearly exhibits what activities are classified under the mining sector and what activities are classified under the mining-related industries sector.

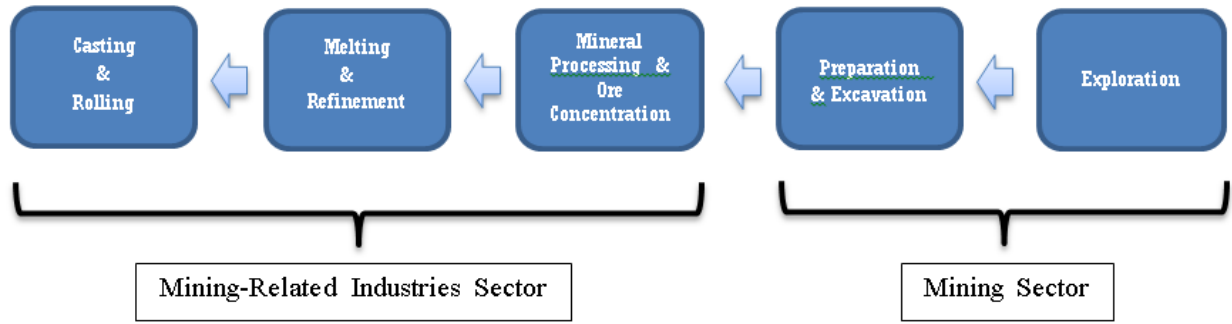

*Figure 1: Separating Activities of the Mining Sector and the Mineral Industry*

With more than 55 billion tons of proven and probable mineral reserves, Iran is one of the top twelve countries in terms of mineral reserves. However, Iran's mining sector has not served Iran's economic development much. Such underground resources have been used by many countries as an opportunity for economic growth and development; however, Iran has not been able to employ these opportunities for a variety of reasons (Mining and Development, 2005). Lack of economic studies with regards to the production process in this economic sector, perhaps, has been one of the missing parts in making use of the mentioned opportunities with Iran's mining sector. Nowadays, economics provides many applied and analytical tools that can be used to study production process deeply. One of these analytical tools is indeed aggregate production function.

The remainder of this paper is organized as follows: After the introduction, the research problems are expressed and discussed. In the first section, the theoretical basis of production functions is examined. The second section reviews some empirical studies in the area of estimating aggregate production functions. In the third section, the role of the mining sector in Iran's economy is analysed. In the fourth and fifth sections, an economic model to estimate the aggregate production function of the mining sector of Iran is specified and estimated, respectively. Finally, the last section summarizes concluding remarks. This paper attempts to answer the following questions:

- How is the structure of production in the mining sector of Iran? Is it capital-intensive or labour-intensive or both?

- What are the effects of technological changes on the output of the mining sector of Iran? Has it caused the output to grow or decline?

## 1. Theoretical Background

A production function is a mathematical equation that represents the relationship between physical inputs and physical outputs. It can represent various ways of combining factors of production to produce goods and services. This function expresses a technical relationship between inputs and outputs in a simple manner. Putting things more accurately, production function is a mathematical equation representing the "maximum" output that a firm can obtain from any fixed and specific set of production inputs at a certain level of technology. Thus, a production function can be represented as an equation in which outcome is considered as the dependent variable and production inputs are regarded as independent variables. Accordingly, the production function can be mathematically expressed as the following equation:

$$Q = f(L, K, ...) \qquad (1)$$

where the dependent variable ($Q$) is the output and the independent variables are various production inputs, such as labour ($L$), capital ($K$), among others. Aggregate production function is a neoclassical economic concept where L>0 and K>0, and is defined as a continuous twice-differentiable function. Its partial derivatives are shown as below:

$$\frac{\partial Q}{\partial L} = f_L \qquad \frac{\partial Q}{\partial K} = f_K \qquad \frac{\partial^2 Q}{\partial L^2} = f_{LL} \qquad \frac{\partial^2 Q}{\partial K^2} = f_{KK} \qquad (2)$$

Furthermore, it is assumed that:

$$f_L \succ 0 \qquad f_K \succ 0 \qquad f_{LL} \prec 0 \qquad f_{KK} \prec 0 \qquad (3)$$

Now, it is guaranteed that the marginal products of inputs are all positive and decreasing.

In any system of production, there are basically two major and distinct concepts in terms of efficiency, one of which is called technical efficiency and the other is called allocative efficiency (Libenstein, 1988). In specifying a production function, it is assumed that engineering and managerial aspects of technical efficiency has previously been considered, so that the analysis can focus on the problem of allocative efficiency. In fact, this is why the correct definition for production function is considered as a relationship between technically "maximum" possible output and the required amount of inputs for producing that output (Shephard, 1970). Despite this, most theoretical and empirical studies define the production function carelessly as a technical relationship between output and inputs, and the assumption that such an output needs to be the "maximum" output (and consequently the minimum inputs) quite often remains unspoken (Mishra, 2007).

Neoclassical aggregate production function is defined on the basis of specific characteristics such as decreasing marginal products of inputs and substitutability of labour and capital with respect to each other. This function can be rewritten in terms of output per-capita and capital per-capita (i.e. per worker in fact) as follows.

$$\frac{Q}{L} = A \cdot f(\frac{K}{L}) \Rightarrow q = A \cdot f(k) \tag{4}$$

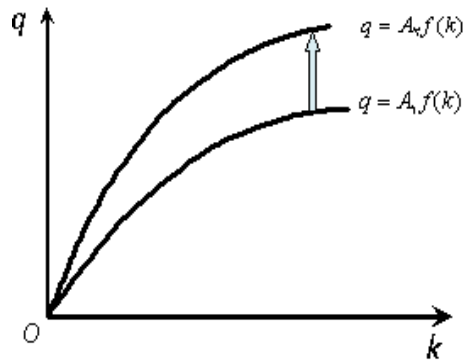

*Figure 2: Neoclassical Aggregate Production Function*

As shown in Figure (2), an increase in the capital input will cause an initial position on the curve to move along the same original production function curve towards higher positions (more output), while an increase in the technology level from $A_1$ to $A_2$ (assuming $A_2 > A_1$) will shift the production function curve upward. This shift means that the same amount of input produces more output. In the economics literature, such a shift is referred to as technological progress[2].

Aggregate production function is a relationship that is used to describe the technical relations among inputs and output at a macro level. Theoretically, this function is the sum of micro production functions. However, there are piles of research studies and papers, which have been conducted from the 1940s onwards, arguing that integrating micro production functions to a macro production function can be quite difficult and problematic (Philip & Fisher, 2003). At least, awareness of these problems seems necessary for every economist who has started empirical research. According to Jonathan Temple (2007), every scholar who intends to specify and estimate aggregate production functions needs to first know these circumstances (Temple, 2007). Sylos Labini believes that it is worth reminding these critiques and criticisms, because a significant and growing number of talented young economists do not know or do not take those criticisms seriously and continues to design and develop various forms of aggregate production functions (Sylos Labini, 1995). This type of problems is classified under the title of "the aggregation problem" in the economics literature. This sort of problem arises partly due to the distinctions between microeconomics and macroeconomics. As long as a production function describes the relationship between output and inputs of a firm, problems will be at the minimum. However, when one attempts to identify the production function of an industry, or a sector, or the economy as a whole, it may become a totally different story. An industry is generally composed of several firms that produce similar or different products. Of course, each of these firms uses inputs in accordance with its costs, returns to scale, production technology, and market conditions. The aggregate production function of an industry is obtained by relating the total quantities of the employed inputs by all firms in the industry to the total products produced by all the firms operating in that industry. These issues and many other issues make the usage of these functions theoretically problematic. It should be noted that the further we move away from a microeconomic level towards a macroeconomic level (i.e. from a firm level to a sub-sector level, and then to a sector level, and finally to the economy level) the more serious the theoretical problems become (Mishra, 2007). In economics, these problems have been long discussed under some topics such as "capital controversy" and "Cambridge-Cambridge debates"[3].

---

[2]. Although such shifts are usually called "technological changes" in the economics literature, it should be noted that these changes do not always occur solely due to "technological" changes. Some economists prefer to use the broader term "technical changes" instead of the term "technological changes". By definition, a "technical change" describes a change in the quantity of output produced from the same quantity of inputs. Hence, a "technical change" does not necessarily have to be "technological", but it also could be "organizational", i.e. a change in institutions, regulations, input prices, and the like. Other economists also use the term "changes in Total Factor Productivity (TFP)" which essentially refers to the same economic concept. After all, I will use the term "technological changes" for this purpose throughout this paper following the primary terminology used in the economics literature.

Nevertheless, empirical evidence suggests that assuming an aggregate production function still allows economic models to fit aggregate data quite well and be highly predictive. As will be discussed in the following sections, evidence has shown that aggregate production functions generally provide a good approximation of real-world phenomena. In a study published on the new growth theories, Temple (1998) concludes that although aggregate production functions are the least convincing component among all the components of the modern macroeconomics, it is still considered by many economists as a prerequisite to understand national income levels and growth rates (Temple, 1998). Therefore, it should be noted that aggregate production functions could not be accused of ad-hoc modelling. According to Temple (2007), critics of aggregate production functions overemphasize the economic theory. As a result, there are still compelling reasons for using these functions in many areas despite the fact that they are still being criticized.

---

[3]. The so-called Cambridge-Cambridge debates are a series of intense theoretical debates occurring during 1950s and 1960s between Joan Robinson and her allies in Cambridge, England on the one hand with Paul Samuelson and Robert Solow in Cambridge, USA on the other hand. In other words, the debate happened between Neo-Ricardians and Neo-Classicals. Harcourt (1969) presents a full report of what happened between them and in their debates.

Now, a brief introduction to production functions will be presented here. Despite the criticisms mentioned above, production functions have had a strong presence in economic studies. They have been widely used in the economic analysis. Here is a list of basic forms of some of these functions.

• Cobb-Douglas production function:

$$Q = A \prod_{i=1}^{n} X_i^{\beta_i} \tag{5}$$

• Production functions with constant elasticity of substitution (CES):

$$Q = A \left[ \sum_{i=1}^{n} \beta_i X_i^{-\rho} \right]^{-\frac{v}{\rho}} \tag{6}$$

where it is assumed that: $\rho \geq 1 \quad \rho \neq 0 \quad v > 0 \quad A > 0$

• Translog production function:

$$\ln Q = \alpha_0 + \sum_{i=1}^{n} \alpha_i \ln x_i + 0.5 \sum_{i=1}^{n} \sum_{j=1}^{n} \beta_{ij} \ln x_i \ln x_j \quad ; \beta_{ij} = \beta_{ji} \tag{7}$$

• Transcendental production function:

$$Q = AK^{\alpha} L^{\beta} e^{\gamma K + \mu L} \tag{8}$$

• Debertin production function:

$$Q = AK^{\alpha} L^{\beta} e^{\gamma K + \mu L + \varepsilon KL} \tag{9}$$

For more information about details and implications of these functions, you can see Zeytoon Nejad (2009).

In this section, theoretical background of production functions in general, and a brief introduction of Cobb-Douglas production function, CES, Translog, Transcendental and Debertin in specific were reviewed. Certainly, many other mathematical functional forms can explain the production process of an economic sector or of the whole economy by limiting real-world features through the adaptation of certain assumptions. To learn about the history of production functions and their path of formation, see Mishra (2007). Reviewing the history of production functions, he provides a brief description of features and characteristics of other production functions such as the VES[4], CMS[5], GPF[6], LINEX[7], multi-output production functions, among many other forms of production functions.

## 2. Empirical Studies

Production functions are of great importance in economics. They are at the heart of the theory of production. They are also applied indirectly in other areas of economics, e.g. in macroeconomics (in both theories of economic growth and business cycles), and also in the investigation of productivity in different economic sectors and at different economic levels. Concerning the estimation of these functions, several studies have been conducted at different economic levels. These studies are reviewed in the following.

Agheli-Kohneshahri (2006) estimated the production function of Iran's mines. He employed the panel data of the mining sector in various provinces of Iran during 1996-2002. To estimate the production functions of Iran's mines, he used logarithmic Cobb-Douglas, Translog and transcendental production functions. Based on the production function estimations conducted by the Pooled Least Square (PLS) and the Generalized Least Square (GLS), he found that Iran's mines have been labour-intensive, and that the return to scale (sum of the elasticities of production with respect to inputs) has been slightly greater than 1. This suggests that there exist increasing returns to scale in the mining sector of Iran.

Nafar (1996) estimated and analysed the production function of Iran's industries to calculate returns to scale and technological progress in these industries. For this purpose, he used cross-sectional time-series data for 1971-1993. In this study, two models of single- and multiple-time trend were used which represented the technical changes of production. Estimation results confirmed a different time trend in the technological progress in production. The results also showed that the single time trend (STT) has a very slow upward trend in the improvement of production technology during the studied period. However, the multiple-time trend (MTT) indicated a very strong reduction in the trend for the year 1971. The results showed diminishing returns to scale in Iran's industries.

---

[4]. Variable Elasticity of Substitution
[5]. Constant Marginal Share
[6]. Generalized Production Function
[7]. Linear Exponential

Lindenberger (2003) estimated the production function for the service sector of Germany. Arguments of the production function involved in his research include technological parameters, labour, capital, and energy. The results indicate that the time average of elasticity of production with respect to capital, labour, and energy have respectively been $\bar{\alpha} = 0.54$, $\bar{\beta} = 0.31$ and $\bar{\gamma} = 0.21$ during 1960-1978 and $\bar{\alpha} = 0.53$, $\bar{\beta} = 0.26$ and $\bar{\gamma} = 0.21$ during 1978-1989.

Shankar and Rao (2012) estimated the long-term growth rates of Singapore. In doing so, they specified a CES production function. The results show that the elasticity of substitution between labour and capital was 0.6; technological progress in the economy of Singapore was labour-augmenting, and long-term economic growth rate was about 1.8%.

Table (1) summarizes the results of the empirical studies discussed above.

*Table 1: Summary of Empirical Studies Regarding the Estimation of Production Functions*

| Author | Country | Time | Level | Data | Inputs | Functions |
|---|---|---|---|---|---|---|
| **Agheli-Kohneshahri (2006)** | Iran | 1996-2002 | Mining sector | Consolidated | K,L | Cobb-Douglas, transcendental and Translog |
| **Nafar (1996)** | Iran | 1971-1993 | Industries | Consolidated | K,L | Translog |
| **Rezapoor & Asefnejad (2005)** | Iran | 1998-2004 | Hospitals | Consolidated | Doctors, nurses, beds, and other staff | Cobb-Douglas |
| **Hadian et al (2007)** | Iran | 2000-2005 | Hospitals | Consolidated | Doctors, nurses, beds, and other staff | Cobb-Douglas |
| **Zeranejad et al (2004)** | Iran | 1979-2002 | Firm | Time series | K,L | Cobb-Douglas, Debertin and Translog |
| **Lindenberger (2003)** | Germany | 1960-1989 | Servicing private sector | Time series | K,L,E | Cobb-Douglas |
| **Antras (2004)** | America | 1998-1948 | | Time series | K,L | Cobb-Douglas |
| **Xiang (2004)** | Canada | 1997-1961 | Macro | Time series | K,L | Cobb-Douglas |
| **Khalil (2004)** | Jordan | 2002 | Industrial manufacturing | Sectional | K,L,M | Translog |
| **Bonga-Bonga (2005)** | South Africa | 2002-1972 | Macro | Time series | K,L | CES |
| **Shankar&Rao (2012)** | Singapore | 2009-1960 | Macro | Time series | K,L | CES |
| **Manonmani (2013)** | India | 2010-1991 | Textile industry | Time series | K,L | Cobb-Douglas |

## 3. Mining Sector in Iran's Economy

Iran's geological formations are mostly located in the Zagros and Alborz mountains and the formation of central plateau is extensively affected by the geological phenomenon of tectonics. Iran's location on the Alps-Himalayan belt has also provided it with huge, diverse mineral resources. There are 62 types of valuable minerals in Iran, which is a rare case throughout the world.

With a vast area, various climates, unique geological and geographical locations, Iran has gained a unique variety of minerals. With more than 55 billion tons of proven and probable mineral reserves (in total), Iran is one of the top twelve countries in the globe. Based on the 2004 statistics and in terms of weight metrics, Iran produces 1.24% of the minerals being produced in the world. Figure (3) depicts the production of minerals by the select countries.

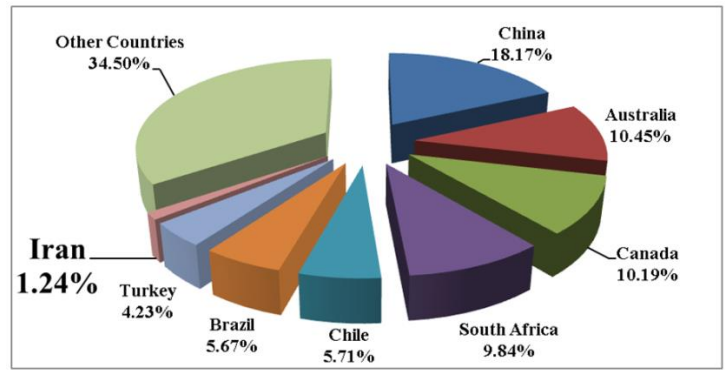

*Figure 3: The Fractions of the Mineral Production of Different Countries in the World (Comparison Done in Terms of Weights)*

However, Iran's mining sector still has not achieved the share it deserves of Iran's economy. Its fraction of GDP is still less than %1. Iran's GDP report in the national accounts during 1976-2006 shows that the average relative share of mining sector in GDP has always been small (almost 0.5%). In contrast, according to the 2004 statistics in, the share of the mining sector in GNP for other countries such as South Africa (7.11%), Australia (6.4%), Denmark (5.4%), Mexico (4.3%) and Canada (3%), have been far larger than that of Iran (Zeytoonnejad, 2005).

Although the share of the mining sector's value added has been small in Iran's economy, it has had an increasing trend during 1976-2006. To investigate this further, see the diagram below.

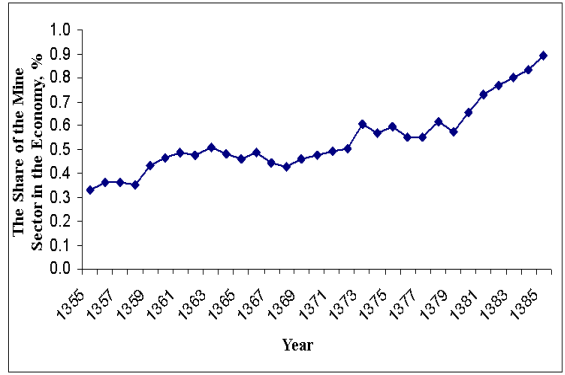

*Figure 4: Time Trend of the Share of the Mining Sector in Iran's GDP*

As indicated in figure (4), the contribution of mining sector to GDP has experienced an increasing trend during 1976-2006.

According to the theory of production, the main variables of production are value added (representing output), labour and capital stock (representing inputs). In this section, we will examine these variables in the period 1976-2006. To this end, time-series diagram of these variables are presented as follows.

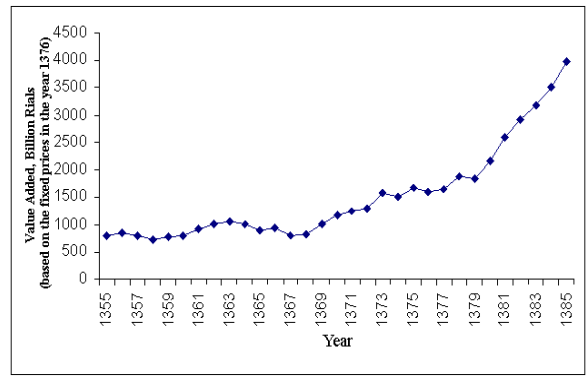

*Figure 5: Value Added of the Mining Sector during 1976-2006*

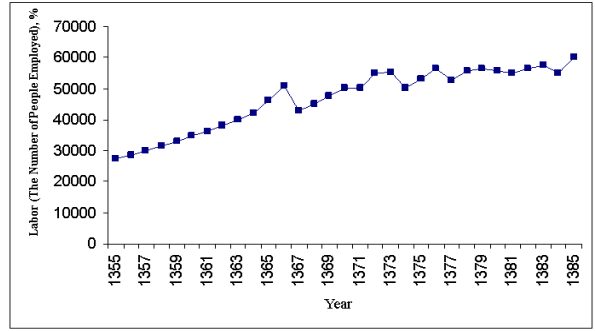

*Figure 6: Employment in the Mining Sector during 1976-2006*

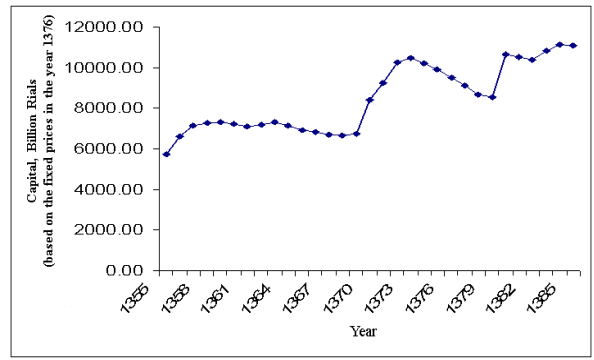

*Figure 7: Time Trend of Capital Stock in the Mining Sector during 1976-2006*

As the above diagrams exhibit, the value added, employment, and capital stock of the mining sector have experienced nearly an increasing trend in the given period. Now, after an overview of the variables influential on the production process, table (2) reports the annual growth rates of these variables.

*Table 2: Average Annual Growth Rates of Value Added, Labour and Capital Stock in the Mining Sector, 1976-2006*

| Variable | Average Annual Growth Rate |
|---|---|
| Value Added | 5.48% |
| Labour | 2.65% |
| Capital Stock | 2.21% |

## 4. The Model and Variables

In this section, the collected data is first introduced. Then, the aggregate production function of the mining sector of Iran is estimated using Cobb-Douglas, transcendental, Debertin and Translog production functions based on the time-series data of 1976-2006.

By definition, value added is the difference between the value of receipts and payments. Value of receipts is indeed the total value of mineral production, saleable mineral waste, construction, major repairs of capital assets by employees, and some other sorts of receipts. In contrast, value of payments is defined as the total value of materials, less-durable tools, consumed fuel, purchased electricity, purchased water, and some other kinds of payments. In this paper, the data acquired from national accounts was used to collect data on the value added of the mining sector. National accounts are published annually by the Central Bank of Iran. This data was considered as the real value added in fixed prices in 1998.

Capital stock refers to total capital goods, which are measured in the same unit. In mining sector, capital goods include durable machinery, vehicles, equipment, buildings and facilities (excluding the land value), special roads for mines, software packages, etc. In other words, different capital goods are converted to a common unit of measure and are summed together. Accordingly, a measure of physical capital stock is obtained in the mining sector. In this study, the data related to the capital stock of Iran's mining sector was taken from estimations of the Central Bank of Iran for the period of 1975-2002. In order to estimate the last 5 years (2003-2007), the estimation method of Central Bank was used as follows.

$$K_t = (1-\delta)K_{t-1} + I_t \qquad (10)$$

where $K_t$, $K_{t-1}$ and $I_t$ are the capital stock in year $t$, capital stock in year $t-1$, investment in year $t$ in terms of fixed prices, respectively, and $\delta$ is the depreciation rate which has been considered 4.7% in accordance with Amini (2006).

By definition, the term labour here refers to all the individuals employed in the mining sector working inside or outside Iran's mines either full-time or part-time. Employees are divided into two categories: production-line employees as well as administrative, financial and service employees. Productive employees are those involved in the exploration, extraction, mineral processing; i.e. those who deal directly with extraction and production, including simple and skilled workers, technicians, engineers and transportation staff. Administrative, financial and service personnel include office, administrative, financial and servicing staff, and also the staff employed at central offices; i.e. those who are not directly involved in the extraction and production processes.

To collect data on labour, this study used Iran's Mines Annual Report, which is published annually by the Statistical Centre of Iran. Although, for this purpose, it was possible to use the estimates made by Iran's Central Bank, Macroeconomics Office and the Office of Planning and Budget Organization, or some other reliable estimates around, finally the data from the Annual Census of Mines was used because of the higher credibility and accuracy of "census data" compared to those of "estimated data". Unfortunately, there were multiple gaps in the census data provided by the Statistical Centre of Iran on the labour employed in the mining sector for years1977-1984, 1991 and 2005.

To fill the data gaps, there are two alternative interpolation methods to choose from, namely the exogenous method and the endogenous method. In this study, the exogenous interpolation method was used to fill the data gaps mainly due to its simplicity. In this method, the data available for the years before and after a gap are used as benchmarks. The exogenous approach assumes the average annual employment growth rate as a constant and estimates the employment rate regardless of changes in variables impacting employment, such as investment and production. Let $m$ represent the average annual employment growth rate between two consecutive periods (two time points) when census has been carried out; then, the employment rate at the time point of year $t$ is obtained as follows:

$$L_t = (1+m)^t \tag{11}$$

In the same way, the few data gaps in the time series of the census data of the mining sector employment which had been collected by the Statistical Centre of Iran were eliminated.

Table (3) reports the used variables and data sources.

*Table 3: Variables Used, Symbols, and Data Sources*

| Symbol | Variable | Resource |
|---|---|---|
| $Q$ | Value Added (Production) | Central Bank of Iran |
| $L$ | Employment | Statistical Center of Iran |
| $K$ | Capital stock | Central Bank of Iran |

As noted earlier, many mathematical functional forms are capable of explaining the production process in an economic sector by limiting the real world through adopting some reasonable assumptions. In order to estimate the production function of the mining sector of Iran, four different types of production function were employed, namely Cobb-Douglas, transcendental, Debertin, and Translog. Table (4) shows the specifications of these functions.

Table 4: A Summary of the Specified Models for Estimation

| Production function | Explained model |
|---|---|
| Unrestricted Cobb-Douglas (C-D) | $\ln Q = \alpha + \beta \ln K + \theta \ln L$ |
| Unrestricted Tinbergen C-D | $\ln Q = \alpha + \beta \ln K + \theta \ln L + \gamma T$ |
| Restricted Per-Capita C-D | $\ln(\frac{Q}{L}) = \ln A + \alpha \cdot \ln(\frac{K}{L})$ |
| Restricted Tinbergen Per-Capita C-D | $\ln(\frac{Q}{L}) = \ln A + \alpha \cdot \ln(\frac{K}{L}) + \gamma T$ |
| Transcendental | $\ln Q = \alpha + \beta_1 L + \beta_2 \ln L + \theta_1 K + \theta_2 \ln K$ |
| Debertin | $\ln Q = \alpha + \beta_1 \ln L + \theta_1 \ln K + \beta_2 L + \theta_2 K + \delta KL$ |
| Translog | $\ln Q = \alpha + \beta_1 \ln L + \beta_2 (\ln L)^2 + \theta_1 \ln K + \theta_2 (\ln K)^2 + \delta (\ln L)(\ln K)$ |

In the production functions mentioned above, coefficients of the trend variables are considered to be factors of technological changes. Thus, a positive sign for these coefficients indicates a technological progress. Conversely, a negative sign indicates a technological deterioration.[8] As shown in the table above, a Cobb-Douglas model can be estimated in multiple possible scenarios. This function was here estimated in restricted and unrestricted forms as well as simple and Tinbergen forms, constituting four cases in total.

## 5. Experimental Results

The results from some of the above models were not consistent with the fundamental theories of economics. Thus, those results are not reported in this paper. In most of these models, the large number of explanatory variables reduces the degrees of freedom, violating the parsimony principle, which may be followed by some other problems. On the other hand, repeating variables in different ways may bring about co-linearity problems for the model, which in turn may influence the standard deviation and statistical significance of the variables. It is noteworthy that a dummy variable for Iran-Iraq war was also entered in the models, and it was found that this war has not had any significant effect in any of the specified models.

Among the specified and estimated models, the unrestricted Tinbergen Cobb-Douglas model provides the best results in compliance with economic theories, which are also consistent with statistical and econometric criteria and principles.

---

[8]. In the theory of economic growth, these changes are usually referred to as "technological progress" vs. "technological deterioration", reflecting the long-term essence of the economic growth theory. In the theory of business cycles, these sorts of changes are usually referred to as "positive technology shocks" vs. "negative technology shocks", reflecting the short-term nature of business cycles. However, here we I use the former terminology as the present study is a long-term study in nature having to do with the theory of economic growth.

Before estimation, the first step is to test the stationarity of variables. Then, the production function of the mining sector is estimated. Next, assumptions and requirements are assessed by their relevant tests.

**Stationarity Test**

Economic analysis assumes that there is a long-term equilibrium relationship between variables considered in an economic theory. In applied econometric analysis to estimate long-run relationships between variables, the mean and variance are assumed to be constant over time and consequently independent of the time factor. Therefore, a behavioural time-consistency is implicitly assumed for the variables being studied. However, empirical research has found that the behavioural consistency of time-series variables is not fulfilled in most cases. Therefore, in these cases, in which time-consistency or so-called stationarity of variables is not fulfilled, the classical t and F statistics resulted from estimation methods are not validated and the results will be misleading. This problem is referred to as "spurious regression". As a result, such variables need to first be tested for stationarity in order to ensure the reliability of results.

A time-series variable is stationary when the mean, variance and autocorrelation coefficients remain constant over time. In other words, if starting time of a series of data is changed but the mean, variance, and covariance remain unchanged, the series will be stationary. A graph of the logged time-series variables employed in this study is presented below.

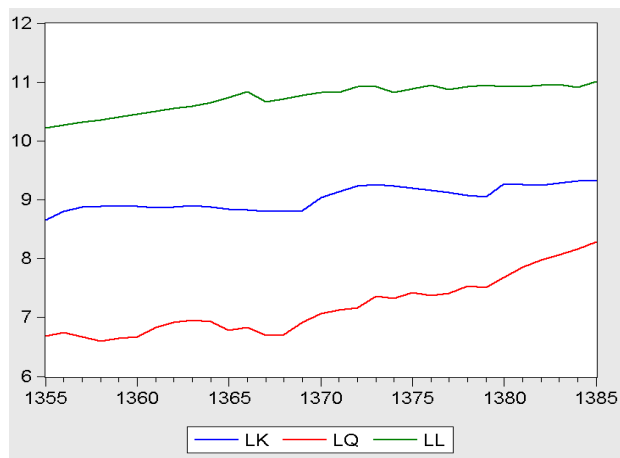

*Figure 8: The Logarithmic Values of the Time-Series Variables Used in the Models*

Figure (8) implicitly suggests the existence of a time trend, and that the above variables follow a time factor. In other words, this graph implicitly indicates the existence of unit roots in the levels of the variables being studied. To test this scientifically, the following tests were taken advantage of in order to be explicit about the stationarity of the above variables.

In particular, two unit root tests, namely Augmented Dickey-Fuller (ADF) and Philips-Perron (PP), were applied for this purpose. The results show that variables are all I(1) [9]. Summary of results is presented in Table (5).

*Table 5: Summary of Results from Reliability Tests Conducted for Logarithm of Variables*

| Variable | ADF level | PP level | First order difference of ADF | First order difference of PP | Result |
|---|---|---|---|---|---|
| LL | -2.157 | -2.071 | -6.257 | -10.904 | I(1) |
| LK | -2.751 | -2.472 | -4.124 | -4.114 | I(1) |
| LQ | -1.120 | -1.120 | -4.979 | -4.984 | I(1) |

In testing the critical values of levels, the critical values at 1%, 5% and 10% were -4.297, -3.568 and -3.218, respectively; for the first order difference, the critical values at 1 %, 5% and 10% were -4.310, -3.574 and -3.222, respectively.

Then, the production function was estimated using the ordinary least-squares estimator. In general, these estimators are the Best Linear Unbiased Estimators (BLUE) according to the Gauss-Markov theorem. However, this method can be used to estimate coefficients when the model satisfies the following assumptions:

- Lack of bias ($E(u_i) = 0$)

- Lack of heteroscedasticity ($E(u_i^2) = \sigma^2 I$)

- Lack of autocorrelation of residuals ($E(u_i, u_j) = 0$ in $i \neq j$

- Lack of correlation between residuals and explanatory variables ($E(X_i u_i) = 0$)

- Normal distribution of residuals with variance $\sigma_u^2$ and the expected value of zero ($u \sim N(0, \sigma_u^2)$)

In estimating the production function of the mining sector, the best model in terms of theoretical principles of econometrics was the unrestricted Tinbergen Cobb-Douglas production function. Considering the conditions and results explained above, we finally decided to estimate this version of Cobb-Douglas production function. Table (6) summarizes the results of the estimation of the aggregate production function of Iran's mining sector:

---

[9]. The same tests for stationarity were conducted on logarithmic values of capital per-capita and value added per-capita; the results of both tests showed that the stationarity was at I(1) for both variables.

*Table 6: Summary of Results from Estimation of Unrestricted Production Function for the Mining Sector*

| Parameters, measures, statistics | Values |
|---|---|
| Elasticity of production w.r.t. Capital (α) | 0.44 |
| *t-Statistic for $\alpha$* | 1.88 |
| Elasticity of Production w.r.t. Labour (β) | 0.41 |
| *t-Statistic for $\beta$* | 1.47 |
| Factor of Technological Change | 0.08 |
| *t-Statistic for the Factor of Technological Change* | 1.67 |
| The Coefficient of AR(1) | 0.9 |
| *t-Statistic for Coefficient of AR(1)* | 10.58 |
| $R^2$ | 98% |
| $\bar{R}^2$ | 97% |
| F-Statistic | 255 |
| Durbin-Watson | 1.93 |
| Number of Observations (n) after Adjustment | 30 |

The results from this estimation show that the elasticity of production with respect to capital and labour has been 0.44 and 0.41, respectively. The coefficient of technological changes was positive and statistically significant in the model, which means a positive impact from technological changes on the output of Iran's mining sector, i.e. a technological progress. To solve the issue of autocorrelation existing in the model, the component AR(1) was added to the model. Due to this addition, the value of Durbin-Watson (DW) approached the numerical value of two, suggesting that there should no longer be any autocorrelation among residuals. The coefficient of determination (R squared) and the adjusted coefficient of determination (adjusted R squared) were 98% and 97%, respectively, indicating the high explainability of the model. In other words, the coefficient $R^2$=98% indicates that 98% of the variations in the dependent variable (value added) can be explained by the explanatory variables (technological changes, labour and capital stock). Statistical significance of coefficients "t" was also satisfactory, so the coefficient of capital (with a 95% confidence interval) and the coefficient of labour and time trend (with a 90% confidence interval) were all statistically significant. The numerical value of F-statistic indicates the significance of the model as a whole.

To see the actual graph of value added and the fitted line (estimated regression line) as well as the plot of error terms, refer to Figure (9). The proper fitness of the regression line on the graph of the value added indicates the suitability of the specified model and the capability of the estimated function in economic forecasts.

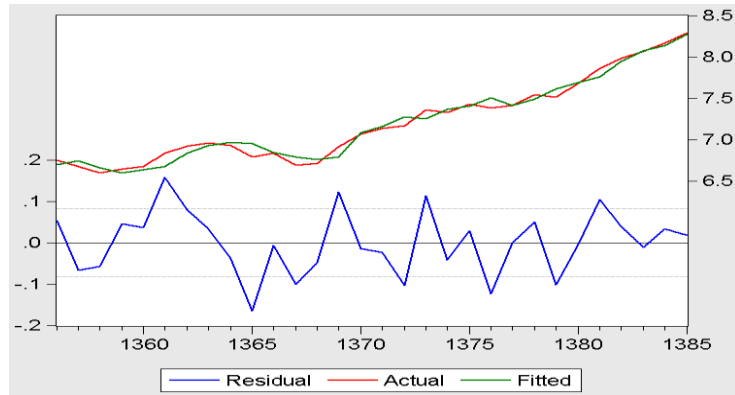

*Figure 9: Fitness of the Estimated Unrestricted Production Function on the Graph of the Value Added together with the Graph of Residuals*

Thus, the characteristics of the estimated model are acceptable according to statistical, theoretical and econometric standards and principles; therefore, the estimated model can be accepted as a reliable model. Now, the model is evaluated for checking the classical assumptions such as heteroscedasticity, normality, etc. The degree of reliability and validity of the model will be determined accordingly.

**Heteroscedasticity Test**

One of the classical assumptions is to have identical variances of residuals (homoscedasticity) in various periods. In other words, $E(u_i^2) = \sigma_u^2 I$ where $i = 1, 2, \ldots , n$. Violation of this assumption brings about a problem called heteroscedasticity. By testing homoscedasticity of variances in the regression through Breusch-Pagan-Godfrey method, it became apparent that the assumption of heteroscedasticity could be rejected. Therefore, there is no problem of heteroscedasticity in the estimated model.

**Autocorrelation test**

To investigate the problem of autocorrelation, Breusch-Godfrey test (LM test) was employed. The results of this test for the original unrestricted model (i.e. prior to incorporating the component AR(1)) implied that there was the problem of autocorrelation in the original estimated model. The diagram of the autocorrelation and partial correlation of the residuals are presented below.

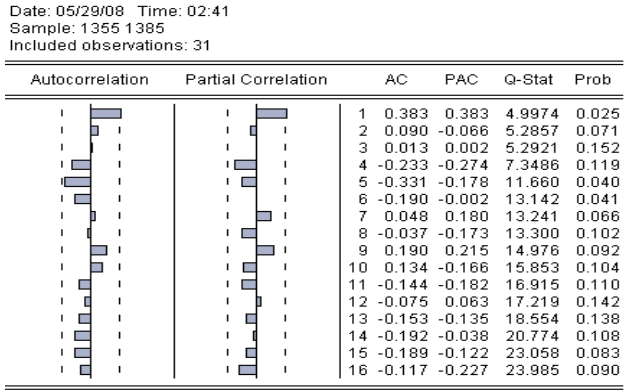

*Figure 10: The Autocorrelation and Partial Correlation of Residuals in the Original Unrestricted Model*

Autocorrelation problem is clearly visible in the graphs above. In order to solve this problem, an AR(1) was added to the model, and thus the problem was resolved. The results of Breusch-Godfrey test, after inserting AR(1) into the model, indicates that the problem has been resolved. The diagram of autocorrelation after the addition of AR(1) component is as the following.

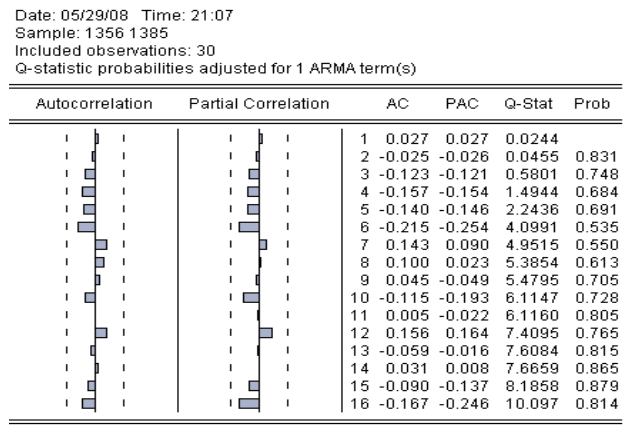

*Figure 11: Autocorrelation and Partial Correlation of Residuals in the Final Unrestricted Model*

Therefore, the addition of the component AR(1) to the model resolved the autocorrelation problem. Now, we will continue to investigate other assumptions of the classical regression model.

**Normality of the Residuals**

The results below show that residuals are distributed roughly normally. To explore this, see figure (12).

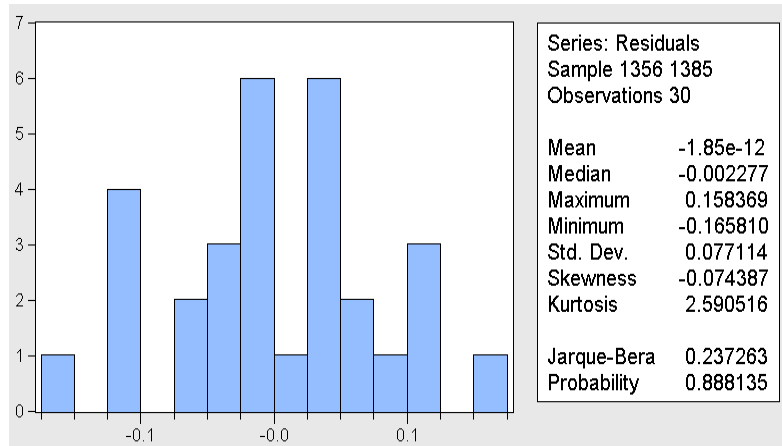

*Figure 12: The Results Obtained from the Statistical Distribution of the Residuals in the Unrestricted Model*

Thus, Figure (12) shows that the assumption of normal distribution for the random process of residuals holds; in other words, $u \sim N(0, \sigma^2)$. This in turn suggests $E(u_i) = 0$ which indicates that the associated classical assumption is also validated. Calculation of $E(u_i)$ shows that its numerical value equals -1/85186E-12, which is almost equal to zero.

**Independence of Residuals from the Explanatory Variables**

Results from calculating the coefficient of correlation between residuals and explanatory variables of capital stock and labour were 0.054 and -0.098, respectively, which suggest the independence of residuals from both explanatory variables in the estimated model. Thus, the assumption of independence of residuals from explanatory variables is validated.

Therefore, it turns out that all of the classical assumptions hold in the estimated model and there is no violation of the underlying classical assumptions.

**Co-linearity**

Co-linearity essentially means a linear relationship and strong correlation between the explanatory variables of the regression model. Even when the co-linearity is too severe, OLS estimators maintain the BLUE features. Some researchers worry about circumstances in which explanatory variables under study are co-linear; nonetheless, co-linearity does not violate any of the underlying assumptions of regression. Even in such cases, unbiased and consistent estimates and errors are correctly made and computed. The only effect of co-linearity is that the estimates of coefficients are obtained with large standard errors. This effect is observed when there are independent variables with small variances in a model. This effect also applies to the times when there are a small number of observations in a model.

Two empirical rules by which the presence or absence of co-linearity can be determined are as follows:

- A high $R^2$ and a small number of significant t-statistics simultaneously in a model
- A high correlation between explanatory variables

The examination of the mentioned empirical rules found no severe co-linearity in the model.

Now, as the final step of the co-integration method, it is necessary to calculate the time series of residuals and test the stationarity of this series. The series of the residuals was obtained using the reports of the computer software Eviews (6). Only if the time series is stationary, the results of ordinary least squares estimation are validated. Using both ADF and PP tests, stationarity of the series was investigated. The summary of the results is listed in Table (7).

*Table 7: The Summary of Results from Stationarity Tests for the Residuals of the Unrestricted Model*

| Variable | ADF level | PP level | Result |
| --- | --- | --- | --- |
| Disturbing elements | -5.096 | -5.093 | I(0) |

The results from both methods suggest that the series is stationary. Thus, the estimated model is acceptable as a reliable model.

## 6. Conclusion

Despite the low contribution of mining sector to Iran's GDP, its direct and indirect effects are considerable on the creation of national wealth. High diversity and abundance of mineral resources in Iran has brought about a considerable potential for Iran's economy to experience faster economic growth and development. Mining sector is of great importance for three major reasons. First and most importantly, it is the main source of providing primary materials needed for running other industries. Second, it can be a primary source of providing employment opportunities. Last but not least, it potentially can play role as a creator of regionally balanced economic development, since most of Iran's mineral resources are situated in or near poor provinces. However, economical exploitation of these mineral resources is possible solely through optimal composition of the primary factors of production, namely labour, machinery, intermediate goods, energy, etc.

By reviewing the theoretical background, the empirical literature and the status of Iran's mining sector, this study estimated the aggregate production function of the mining sector and revealed that the structure of Iran's mining sector is both labour-intensive and capital-intensive.

This is due to the fact that the elasticity of production with respect to capital and labour has been 0.44 and 0.41, respectively, which are not considerably different. On the other hand, the production has been operating in the economic zone with respect to both inputs. The coefficient of returns to scale has been 0.85 for this sector, indicating a decreasing return to scale. Moreover, the coefficient of time trend, as the index of technological changes in production over time, was significant, suggesting a positive effect for technological change on the output quantity of the mining sector.

Therefore, promoting the level of technology, creating stability in order to provide a bed for investment, developing infrastructures, making economic policies to increase incentives, increasing R&D, revising scales of production in order to use economies of scale, and developing information banks of Geology and exploration can be helpful in the improvement of efficiency and the increase of output quantity in this sector.

**Resources**